\documentclass{appolb}

\def\be{\begin{equation}}
\def\ee{\end{equation}}
\def\lsim{\raise0.3ex\hbox{$<$\kern-0.75em\raise-1.1ex\hbox{$\sim$}}}
\def\gsim{\raise0.3ex\hbox{$>$\kern-0.75em\raise-1.1ex\hbox{$\sim$}}}

\begin{document}
\pagestyle{plain}
\newcount\eLiNe\eLiNe=\inputlineno\advance\eLiNe by -1
\title{THE ENERGY DEPENDENCE OF THE SATURATION SCALE
\thanks{Presented by D. Schildknecht at Photon05, Warsaw, 31.08. 
to 04.09.2005\\
email: Dieter.Schildknecht@physik.uni-bielefeld.de}%
}
\author{Masaaki KURODA
\address{Institute of Physics, Meiji-Gakuin University, Yokohama 244, Japan}
\and
Dieter SCHILDKNECHT
\address{Fakult{\"a}t f{\"u}r Physik, Universit{\"a}t Bielefeld,
Universit{\"a}tsstrasse 25, 33615 Bielefeld, Germany}}
\maketitle

\begin{abstract}
At low $x \cong Q^2/W^2 << 1$, in deep inelastic scattering, the photon
fluctuates into a $q \bar q$ vector state that interacts via two
gluons with the proton. The energy dependence is determined by the
saturation scale, in our approach given by $\Lambda^2_{sat} (W^2) \sim
(W^2)^{C_2}$. Imposing DGLAP evolution, we find $C_2^{theory} = 0.27$ in
agreement with the model-independent analysis of the HERA data. Different
values of the exponent $C_2$ are correlated with different ratios of the
longitudinal to transverse structure function. This stresses the need for
experiments to separate longitudinal and transverse contributions in
deep inelastic scattering.
\end{abstract}


In deep inelastic scattering (DIS) at low $x \cong Q^2/W^2 < 0.1$ the
photon fluctuates into on-shell quark-antiquark vector states 
(e.g. \cite{Photon})
which interact with the proton via coupling of two gluons \cite{Low} to
the $q \bar q$ color dipole \cite{NZ}. The cross section depends on the 
effective transverse three momentum of the gluon, $\vec l_\bot$, 
absorbed by the 
$q \bar q$ pair in the imaginary part of the virtual-photon forward 
Compton amplitude . The effective transverse momentum of the absorbed
gluon gives rise to a novel scale, the ``saturation scale'', characteristic
for DIS in the $x \to 0$ limit. Since the photon fluctuates into an on-shell
$q \bar q$ pair, the $\gamma^* p$ energy, $W$, besides $Q^2$, is the 
appropriate variable for the virtual photon-absorption cross section to 
depend on. The color-dipole cross section, $\sigma_{q \bar q p} 
(\vec r, W^2)$, and the saturation scale, 
$\Lambda^2_{sat} (W^2)$, become functions of $W^2$.

In the high-energy limit, with 
\be
\Lambda^2_{sat} (W^2) = \frac{1}{6} \langle \vec l^{~2}_\bot \rangle
= \frac{1}{6} B^\prime \left( \frac{W^2}{1 GeV^2} \right)^{C_2},
\label{1}
\ee
the fit to the DIS data gave \cite{Surrow}
\begin{eqnarray}
C^{exp}_2 & = & 0.27 \pm 0.01,\nonumber \\
B^\prime  & = & 0.340 \pm 0.063 GeV^2.
\label{2}
\end{eqnarray}
In terms of $\Lambda^2_{sat} (W^2)$, the cross section of this QCD-based
generalized vector dominance - color dipole picture (GVD-CDP) in
the limit of $Q^2 \to 0$ and $Q^2 \to \infty$ reads
$$
\sigma_{\gamma^*p} (W^2, Q^2) = \frac{\alpha}{3 \pi} R_{e^+e^-} 
\sigma^{(\infty)} \cases{ln \frac{\Lambda^2_{sat} (W^2)}{Q^2 + m^2_0},
& $(Q^2 << \Lambda^2_{sat} (W^2)),$ \cr
\frac{\Lambda^2_{sat} (W^2)}{2 Q^2}, & $(Q^2 >> \Lambda^2_{sat} (W^2))$.\cr
}\eqno (3)
$$
The cross section depends on the single scaling variable \cite{Diff2000,Surrow}
\be
\setcounter{equation}{4}
\eta \equiv \frac{Q^2 + m^2_0}{\Lambda^2_{sat}(W^2)},
\label{4}
\ee
and at HERA, $\Lambda^2_{sat} (W^2)$ varies in the range of 
$2 GeV^2 \lsim \Lambda^2_{sat} (W^2) \lsim 7 GeV^2$. In (3),
$\sigma^{(\infty)} = 27.5 mb$ and $R_{e^+e^-} = 3 \sum_f Q^2_f = 10/3$.
The cross section (3), at any fixed $Q^2 >> \Lambda^2_{sat} (W^2)$ contains
the spectacular strong rise of the structure function
\be
F_2 (x,Q^2) = \frac{Q^2}{4 \pi^2 \alpha} \sigma_{\gamma^*p} (\eta (W^2, Q^2))
\label{5}
\ee
with energy at HERA \cite{HERA}. For $\Lambda^2 (W^2) >> Q^2$, the
saturation limit of weak energy dependence, as seen in photoproduction,
will be reached . Except for exceedingly small $Q^2$, this only happens
at energies far beyond the HERA energy range.

A fundamental question concerns the magnitude of the exponent $C_2$
in (1) that determines the rise of the cross section at large $Q^2$. It
turned out \cite{Ku-Schi} that an examination of the expression for the proton structure
function (5) in the large-$Q^2$ limit in (3) in terms of the dual language
of sea-quark and gluon distributions, upon applying DGLAP 
evolution \cite{Prytz}, provides
an answer to this very important question on the magnitude of $C_2$.

The structure function in the large-$Q^2$ limit of (3) takes the 
form \cite{Ku-Schi}
\be
F_2 (x, Q^2) = \frac{R_{e^+e^-}}{36 \pi^2} (T (W^2) + \frac{1}{2} L (W^2)),
\label{6}
\ee
where $T(W^2)$ and $L(W^2)$ are integrals over the first moments of
the gluon transverse momentum formed with the dipole cross sections in
momentum space. The longitudinal part of $F_2 (x, Q^2)$ in (6) is
expressible in terms of the saturation scale,
\be
L (W^2) = \frac{\sigma^{(\infty)}}{\pi} \Lambda^2_{sat} (W^2).
\label{7}
\ee
The (successful) representation of the experimental results according
to (3) was based on the assumption of $T(W^2) = L (W^2)$ in (6), i.e.
equal magnitude of transverse and longitudinal $(q \bar q)^{J=1}$
scattering cross sections.

We turn to the sea quark and gluon distributions corresponding
to the structure function (6) and the saturation scale in (7).
The structure function measures the sea-quark distribution,
\be
(qq^-)_{sea} \equiv x \Sigma (x, Q^2) = \frac{1}{3 \pi^2} (T (W^2) +
\frac{1}{2} L (W^2)),
\label{8}
\ee
and the gluon distribution is related to the longitudinal part of $F_2$,
i.e. with (7)
\be
\alpha_s (Q^2) x g (x, Q^2) = \frac{1}{8 \pi} L (W^2 = Q^2/x) = \frac{1}{8 \pi}
\frac{\sigma^{(\infty)}}{\pi} \Lambda^2_{sat} (W^2).
\label{9}
\ee

So far, our results have been based on a general analysis of the generic
structure of the two-gluon exchange from QCD. We adopt the assumption that
the sea-quark and gluon distributions in good approximation
have identical dependence on the 
kinematic variables, in our case $W^2 \cong Q^2/x$. They are assumed to
be proportional to each other,
\be
x \Sigma (x, Q^2) = \frac{8}{3 \pi} (r + \frac{1}{2}) \alpha_s (Q^2)
x g (x, Q^2),
\label{10}
\ee
with $r = const \ge 0$. Substitution of the sea-quark and the gluon
distribution from (8) and (9) into (10) yields
\be
T (W^2) = r L(W^2) 
\label{11}
\ee
and $F_2$ in (6) becomes (with (7)),
$$
F_2 (x, Q^2) = \frac{R_{e^+e^-}}{36 \pi^2} T (W^2) (1 + \frac{1}{2r})
\cong \cases{T, & $(r>>1)$,\cr
\frac{3}{2} T, & $(r = 1)$, \cr
L, & $(r \to 0)$. \cr }
\eqno (12)
$$
The above-mentioned (successful) representation of the experimental data
corresponds to $r = 1$.

In the range of sufficiently large $Q^2$, we are concerned with, the
evolution of $F_2 (x, Q^2)$ is in very good approximation determined by
the gluon structure function alone, \cite{Prytz}
\be
\setcounter{equation}{13}
\frac{\partial F_2 (\frac{x}{2}, Q^2)}{\partial ln Q^2} = 
\frac{R_{e^+e^-}}{9 \pi} \alpha_s (Q^2) x g (x, Q^2).
\ee
Upon substitution of (12) and (9), with (11), the evolution in 
$Q^2$ is converted into a derivative with respect to $W^2$ that
determines the $W^2$ dependence,
\be
(2 r+1) \frac{\partial}{\partial ln W^2} \Lambda^2_{sat} (2 W^2) =
\Lambda^2_{sat} (W^2).
\label{14}
\ee
Substituting the power law (1), into (14), we obtain
\be
(2r+1) 2^{C_2} C_2 = 1,
\label{15}
\ee
or
$$
C_2^{theor} = \cases{0, & $(r >> 1)$,\cr
0.276, & $(r = 1)$, \cr
0.65,  & $(r = 0)$.\cr}
\eqno (16)
$$
According to (16), the exponent $C_2$ determining the rise of 
$\sigma_{\gamma^*p} (W^2, Q^2)$ at fixed $Q^2 >> \Lambda^2_{sat} (W^2)$,
or, equivalently, the rise of $F_2 (x, Q^2)$ as a function of $x = Q^2/W^2$
at fixed $Q^2$, is uniquely connected with the relative  magnitude of
sea versus gluon distributions. For a given sea distribution, a very small
gluon distribution ($r >> 1$ according to (10)) and a purely transverse 
structure function (according to (12)) implies a very weak energy
dependence. In contrast, in the limit of $r \to 0$, where the relative
magnitude of gluon-to sea distribution is maximal, we must have an 
extraordinarily strong energy dependence, $C_2 = 0.65$. The previous fit
to the experimental data was based on $r = 1$. The result of the fit 
given in (2)
is in excellent agreement with the theoretical result (16).

The foregoing intimate connection between a strong rise of $F_2 (x,Q^2)$ with
decreasing $x$ at fixed $Q^2$ and a large longitudinal photoabsorption
cross section, and a weak rise for a small longitudinal contribution,
can be further illuminated by measurements allowing to separate
longitudinal and transverse cross sections. Such measurements are in fact
urgently needed.
\vspace*{1cm}

\leftline{\it Acknowledgement}

This work was supported by Deutsche Forschungsgemeinschaft under
grant No. Schi 189/6-1.

\end{document}